# Future Hadron Physics at Fermilab

Jeffrey A. Appel

*Fermi National Accelerator Laboratory*
*PO Box 500, Batavia, IL  60510 USA*
*appel@fnal.gov*

**Abstract.** Today, hadron physics research occurs at Fermilab as parts of broader experimental programs. This is very likely to be the case in the future. Thus, much of this presentation focuses on our vision of that future – a future aimed at making Fermilab the host laboratory for the International Linear Collider (ILC). Given the uncertainties associated with the ILC - the level of needed R&D, the ILC costs, and the timing - Fermilab is also preparing for other program choices. I will describe these latter efforts, efforts focused on a Proton Driver to increase the numbers of protons available for experiments.  As examples of the hadron physics which will be coming from Fermilab, I summarize three experiments: MIPP/E907 which is running currently, and MINERνA and Drell-Yan/E906 which are scheduled for future running periods. Hadron physics coming from the Tevatron Collider program will be summarized by Arthur Maciel in another talk at Hadron05.

## INTRODUCTION

In order to envisage the future of hadron physics at Fermilab, it will be useful to know the research context at the Laboratory.  That can be best done by looking at the recent presentations by Fermilab's new Director, Piermaria Oddone, who began his tenure as Director on July 1, 2005.  Pier has had several opportunities to project the vision of Fermilab's future, both inside the Laboratory and to outside committees.[1] I draw heavily on his presentations.

## THE VISION OF FERMILAB'S FUTURE

The vision of Fermilab's future is driven by physics; in particular, where we see the largest mysteries. We know next to nothing about the next accessible energy scale. This is demonstrated by the plethora of models for what we might find once physics results are available from the Large Hadron Collider (LHC), the machine which will allow access beyond that at the Fermilab Tevatron. We know very little about the world of neutrinos, having only just discovered that they have mass and not knowing even the scale of all the mixing in the neutrino sector. We are ignorant also about what dark matter and dark energy are.

We have strong clues that the areas where these mysteries reside are the hunting grounds for major new discoveries, discoveries which will change our most basic understanding of nature. We also believe there is a deep connection among all these areas.

In addition to the physics context driving the vision of Fermilab's future, there is a strategic context, the U.S. contributions to frontier physics research. If we examine the US domestic accelerator program in 2005, 2010, and 2015, we see the US playing a leading role today in the neutrino frontier, the flavor frontier, and the energy frontier. However, with no new investment, we will be playing a secondary role in all these areas by 2010, and beyond. With investments of new and redirected resources, we can continue to play a leading role in neutrino physics in 2010 and beyond, and return to a leading role at the energy frontier in 2015. We do not presently see how we can play more than a secondary role in flavor physics in the mid-term future, given the plans to turn off the Tevatron and PEPII around the end of this decade.

## Highest Priority New Initiative – The ILC

New initiatives at Fermilab will be consistent with our "overarching goals." We want to enable the most powerful attack on the fundamental science questions of our time by providing world class facilities for HEP as part of the global network. We will develop both the science and technology for particle physics and cosmology research. Our specific goals are to make vital contributions to particle physics in the next decade by developing a powerful new tool for discovery at the energy frontier (the International Linear Collider, ILC) and, at the same time, maintaining Fermilab's foremost accelerator-based neutrino program (including the off-axis experiment NOvA, developing a more powerful proton driver, and more).

The energy frontier for Fermilab already includes our efforts at the LHC as the host laboratory for the US LHC Accelerator Research Program and the US CMS collaboration. However, our goal is to also to play a leading role for the ILC. In collaboration with our partners around the world organized via the Global Design Effort (the GDE for short), we will establish all the technical components, costs, engineering designs, and management structures to enable an "early" decision (by 2010) on building the ILC if that is the right decision at that time. We will position the US (and Fermilab) to host the ILC. We will work to position the US (and Fermilab) to play major roles in detector development and physics analysis at the ILC. This is the highest priority initiative for the laboratory.

The ILC research and development roadmap includes milestones for pressing for the early decision on building this energy frontier facility at or near Fermilab. Such a decision might be as early as 2010. On the way, we will see superconducting radio-frequency module assembly and testing, a GDE Reference Design Report (RDR) effort, ILC component industrialization – all with strong Fermilab contributions. Given an LHC discovery by 2010, the scene will be set for a "GO" signal by the science community and by governments. On the other hand, if it is clear from the GDE RDR either that an ILC is not affordable at that point or that a much longer R&D effort is needed, other program choices will be induced. If there is no discovery at the LHC by 2009, again, there will be other program choices to be made. Today, Fermilab's major effort preparing for other program choices focuses on a Proton Driver, an option that would expand our neutrino physics capability substantially.

# "Ships of the Line"

While we develop the ILC, we must deliver on our "ships of the line". These include our collider programs, our neutrino efforts, and other elements of our diverse program. Our collider program includes experiments at both the Tevatron, where we have the CDF and DZero experiments, and at the LHC where we play an important role on CMS. Our current neutrino program includes MINOS, which is receiving neutrinos produced by protons from the Fermilab Main Injector, and MiniBooNE, which uses neutrinos coming from our "Booster neutrino beam". While focusing on these "ships of the line", we will maintain our scientific vitality with a diverse program that includes smaller accelerator-based experiments, particle astrophysics, theory, and computing and other technology development.

The Tevatron program provides our greatest window into new phenomena until the LHC comes on-line. Between CDF and DZero, there are 1500 collaborators including 600 students and postdocs. The continued success of this program is critically dependent on an ever increasing rate of luminosity integration. The doubling time for luminosity will be a major consideration. This talk will leave a description of the hadron physics at the Tevatron to Arthur Maciel, who will speak on "Hadron Physics at the Tevatron".

Fermilab is the only major US laboratory associated with the CMS experiment at the LHC. As such, Fermilab has a central support role for the US community. This has been the case during construction. Attention is now turning to the huge data and physics discovery challenges. We are establishing an LHC Physics Center (LPC) at Fermilab to coordinate and support these efforts.

The present neutrino program includes two running experiments: MiniBooNE and MINOS. They operate with beams of different energy and with detectors placed at appropriate distances to investigate separate neutrino-mass-difference scales. MiniBooNE has been operating since 2003, and should have its first oscillation results this calendar year. It is looking for $\nu_e$ appearance in a $\nu_\mu$ beam. As Oddone has said, "All hell will break loose if [there is a] positive signal." At a minimum, the conventional picture of three neutrino generations will disappear. The MINOS program is just starting. It will make the most precise determination of the atmospheric neutrino mass difference by measuring $\nu_\mu$ disappearance.

| TABLE 1. Fermilab's Currently Running Neutrino Experiments. | | |
|---|---|---|
| **Experiment** | MiniBooNE | MINOS |
| **Peak Neutrino Energy** | 1 GeV neutrinos | 2 GeV neutrinos |
| **Proton Source Used** | Booster | Main Injector |
| **Detector(s)** | 800 ton oil Cerenkov detector at 0.54 km | 5.4 Kiloton far detector at 735 km and 1.0 Kiloton near detector |

## New Initiatives in Neutrino Physics

We need to measure the full neutrino mixing matrix, and understand it in relation to the quark mixing matrix. However, we do not yet have more than a limit on the value of $\sin^2(2\theta_{13})$. We do not know the neutrino mass hierarchy; i.e., the mass ordering of the mass eigenstates, much less their actual values. We do not know if there is CP violation in the neutrino sector as there is in the quark sector, much less the value of the phase parameter δ. Fermilab is in the best position to make vital contributions to resolve these issues. However, we will need new initiatives in the neutrino program to address the issues. Today, Fermilab has the most powerful beam in the world, at 200 KW, dedicated to the production of neutrinos. Upgrades by 2008 should give 440 KW of primary proton power. When we stop the Tevatron, we'll have more than 600 KW, the same as JPARC will have some time after 2010. We are looking at further improvements; aimed at 1-2 MW of proton beam power.

Our most ambitious efforts to increase the reach of our neutrino program are aimed at developing a Proton Driver, a new source of 8 GeV protons to transfer into our Main Injector ring. We are focused on superconducting rf technology, making use of the R&D for the ILC. The Proton Driver would be a linear accelerator, too, providing a base for extremely flexible operations and a much simplified complex. The Proton Driver energy is only about 2% of an initial ILC. So, proceeding with the Proton Driver could establish useful industrial capability for a full ILC, not to mention provide useful operating experience. Nevertheless, we are also developing alternatives to a full Proton Driver project. Not running the Tevatron means we have the Recycler, Accumulator, and Debuncher rings to think about using. We have VERY EARLY creative ideas on how to take advantage of these opportunities to deliver 1-2 MW at 120 GeV.

## Interlinked Roadmap

The interlinked roadmap just described establishes the main trunk and branches for the laboratory. It supports a vital role for the US community in the global program. However, it does require "strong focusing" now, which narrows the program. On the other hand, it establishes the strong base on which breadth and "texture" can flourish as we move forward.

## HADRON PHYSICS NOW AND IN THE NEAR TERM

Today, hadron physics at Fermilab occurs as part of broader programs at the laboratory. This is evident from Table 2, which lists the three components of Fermilab's accelerator-based experiments. We anticipate adding NOνA and MINERνA, two neutrino-beam experiments, and a Drell-Yan fixed-target experiment, E906, in coming running periods. The current draft long range schedule (Figure 1) shows these initiatives, as well as continuing commitments to the operation of test beams, and several "OPEN" periods where there are currently no commitments.

| Table 2. Fermilab's Accelerator-Based Experiments. | |
|---|---|
| Hadron Collider Experiments | CDF and DZero at the Tevatron <br> CMS at the LHC |
| Neutrino Experiments | MiniBooNE and NuMI/MINOS now <br> NuMI/MINERvA, NuMI/NOvA |
| Fixed Target - SY120 | MIPP/E907 now, Drell-Yan/E906 later, kaons? <br> Test beams |

**Figure 1.** Fermilab Draft Long Range Accelerator Experiment Schedule.

Three experiments are particularly relevant to Fermilab's current hadron physics efforts: MIPP/E907, Drell-Yan/E906, and MINERvA. I will discuss each of these in turn.

# Main Injector Particle Production (MIPP/E907)

Using primary protons from the Main Injector, MIPP/E907 is currently taking data on particle production, both by primary protons at 120 GeV/c, and with tagged $\pi^{\pm}$, $K^{\pm}$, and $p^{\pm}$ beams from 5 GeV/c to 90 GeV/c. These beams impinge on a variety of targets: liquid hydrogen and solid targets of various nuclei. The downstream spectrometer is well equipped to measure multiparticle final states, even states with high multiplicity. The detector includes a large Time Projection Chamber, a Ring Imaging Cerenkov counter, two large-aperture magnets, tracking wire chambers, and calorimeters.

The MIPP physics program begins with testing a proposed scaling law of particle fragmentation. This scaling states that the ratio of a semi-inclusive to an inclusive cross section is a function of the invariant mass $M^2$ only.[2]

$$\frac{\sigma(a+b \to c + X_{subset})}{\sigma(a+b \to c + X)} \equiv \frac{\sigma_{subset}(M^2, s, t)}{\sigma(M^2, s, t)} = \beta_{subset}(M^2)$$

European Hybrid Spectrometer data confirms this scaling in 12 reactions, but only at fixed s.[3] MIPP will test this scaling in 36 reactions, as a functions of both s and t, and for various particle types a, b, and c. The scaling law is said to follow naturally if the reaction factorizes into a two step process: (1) formation of a pseudo-resonance and (2) decay of the pseudo-resonance into the final state.

A second part of the physics program of MIPP is to understand, in detail, the particle production model going into Monte Carlo simulation programs. Date will be taken on thin targets and on the actual NuMI thick graphite target. The flux of neutrinos produced in this NuMI target depends directly on the distribution of hadrons leaving the target. MIPP took particle production data with the NuMI target in June and July of this year. Analysis and use of this data should lead to a significant reduction in the systematic error in the comparison of neutrino events in the MINOS near and far detectors, critical to the neutrino oscillation measurements of MINOS. There is a growing appreciation today of this aspect of systematic errors. One may ask whether the current MIPP measurements will be sufficient for future experiments; e.g., NOvA and T2K.

## Drell-Yan Experiment E906

The Drell-Yan Experiment E906 is an extension of the previous fixed-target experiment E866 – exploring the sea quarks of the proton, the ratio of down antiquarks to the up antiquarks. The detector and beam energy define the acceptance range in $x_{target}$ and $x_{beam}$. In fixed target experiments such as E866 and E906,

$$x_F = x_{beam} - x_{target} > 0.$$

The high-x beam partons are essentially only valence quarks. The low- and intermediate-x target sea partons are selected by the Drell-Yan process. These are the partons which can annihilate with the beam valence quarks to produce the high-mass di-muons of the Drell-Yan process. Thus, the fixed-target Drell-Yan process probes target antiquarks.

One may ask whether the proton sea is isospin symmetric. Furthermore, one may ask by what process is the proton sea formed? If the sea arises from perturbative QCD, then

$$\bar{d}(x_t) \equiv \bar{u}(x_t)$$

The study of the ratio of cross sections for deuterium to hydrogen gives the ratio

$$\left.\frac{\sigma^{pd}}{2\sigma^{pp}}\right|_{x_b \gg x_t} \approx \frac{1}{2}\left[1 + \frac{\bar{d}(x_t)}{\bar{u}(x_t)}\right].$$

E866 data clearly show an asymmetry. E906 will extend the measurements to higher x as shown in Figure 2.

Another aspect of the E906 physics program is to look for the effects of nuclear pions, an expected antiquark enhancement in nuclei relative to nucleons. Early predictions (Berger and Coester[4]) were proven false by Fermilab E772 Drell-Yan data. However, the E772 data has relatively large uncertainties, especially as x increases. E906 will challenge the revised predictions, shown in Figure 3. The new models for the ratio of cross sections on iron and hydrogen shown there still diverge sharply from unity at larger Bjorken-x – and from each other.

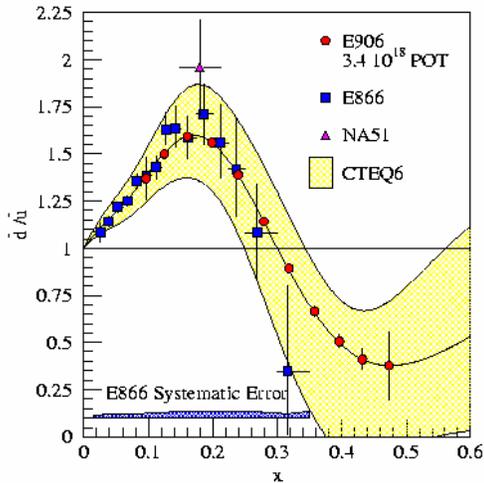
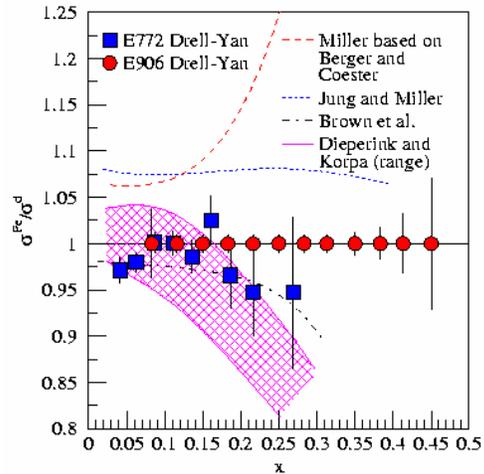

**FIGURE 2.** Bjorken-x dependence of existing and projected data on the ratio of down antiquarks to the up antiquarks in the proton sea.

**FIGURE 3.** Drell-Yan cross section ratio on iron relative to deuterium, showing predictions[4] for the effects of nuclear pions.

## MINERνA

MINERνA is a low-energy neutrino-nucleus scattering experiment using the NuMI beam. The name of the MINERνA experiment derives from the capital letters in "Main INjector ExpeRiment ν−A". MINERνA's main physics objective is to make measurements of the neutrino-nucleus scattering processes important for minimizing the systematic errors in neutrino oscillation experiments such as the cross sections for quasi-elastic, resonance and coherent π production. MINERνA will also significantly augment studies by the nuclear physics community at Jefferson Lab such as using the weak current to study the transition region of perturbative to non-perturbative QCD and the use of a variety of nuclear targets (Pb, Fe and C) to investigate nuclear effects in ν-induced interactions a topic of considerable importance for oscillation experiments as well.

In order to distinguish among the various processes, the detector has a high-granularity, fully-active scintillator-strip central core surrounded by electro-magnetic and hadron calorimeters. Triangular-shaped scintillator extrusions yield position precision through light-sharing between adjacent strips. The MINERvA detector will be placed just upstream of MINOS near detector.

A statistical sample of over 16 million charged current events will be collected in a four-year run. The experiment received Stage I approval in April, 2004, and had a successful Director's Review in January, 2005. The funding is anticipated to come from the Department of Energy through Fermilab, with construction and installation complete in the fall of 2008; physics data-taking starting in 2009.

In addition to the oscillation-motivated measurements, MINERvA will measure the poorly known high-x parton distribution functions, strange particle production channels crucial for proton decay experiments and the recently developed generalized parton distribution functions (GPDs) via the weak deeply-virtual Compton scattering process, shown in Figure 4. This will be the first measurement of GPDs with neutrinos, and as such, will allow flavor separation of GPDs. According to a calculation by A. Psaker,[5] MINERvA would accumulate 10,000 weak deeply-virtual Compton scattering events in a 4-year run.

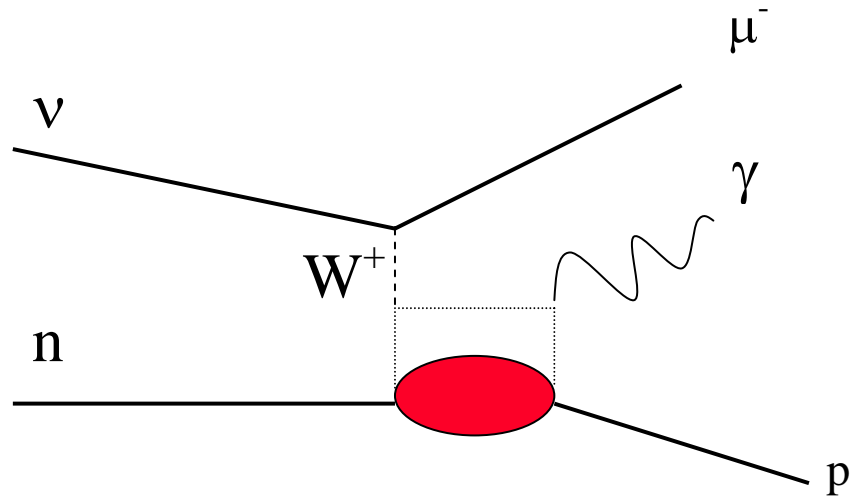

**FIGURE 4.** Weak deeply-virtual Compton scattering, an exclusive reaction with small momentum transfer t and a high energy photon γ.

## A PROTON DRIVER IN FERMILAB'S FUTURE?

There is a need for more intense proton sources for future long-baseline neutrino-physics experiments. This couples to a need to replace the "old" upstream elements in the Fermilab accelerator chain, the Linac and Booster. A common solution to these needs, as mentioned in the discussion of Fermilab's future, would be to build a new proton source to replace the existing Linac and Booster, a Proton Driver.

## The Proton Driver Parameters and Technology

The high-level parameters discussed for a Fermilab Proton Driver are listed in Table 3. The favored implementation for the Proton Driver is a superconducting radio-frequency module linac. It has several advantages over a synchrotron solution: (1) better performance, (2) more flexibility of beam types, and (3) a direct connection to the recently announced "cold technology" choice for the ILC. Thus, the development of the technology, any industrialization of the manufacture of components for the Proton Driver, and any operating experience can feed directly into ILC efforts. It is clear how this fits into the vision of the ILC timeline and of Fermilab's future.

| Table 3. High-Level Parameters for a Fermilab Proton Driver. | |
|---|---|
| Beam power at 8 GeV | 0.5-2.0 MW |
| Beam power at 120 GeV | 2.0 MW |
| Resulting increase in beam power of the Main Injector beam. | 10 x today's power |

## The Physics Case for the Proton Driver

The Physics case for a Proton Driver has been documented[6] following a workshop (October 6-9, 2004) and study[7] chaired by Steve Geer. The study and workshop summarized results from several working groups, including those on (1) neutrino interactions, (2) kaons and pions, (3) neutrons, and (4) antiprotons. Much of the physics considered is hadron physics.

## FINAL COMMENTS

The DOE has announced that Fermilab will be the home of accelerator-based HEP in the US in the future, even after the Tevatron is turned off. Thus, defining the future of much of HEP hadron physics in the US and beyond will depend on a combination of funding and user demand for facilities - two tightly coupled things.

I have summarized a few dedicated hadron physics experiments which already fit into other, directly funded programs (Tevatron and neutrinos, but also a commitment to a test beam for ILC and other detector R&D). Being part of other programs may be the best model for future hadron physics at Fermilab.

What will be the total available beam? Will we build a Proton Driver? What fraction of the available protons will go to hadron physics? Will a stretcher ring be available (either using the Tevatron or a new ring to be proposed and built?

A robust program depends on robust interest, and eventually, compelling physics-driven experiment proposals.

## ACKNOWLEDGMENTS

I would like to acknowledge the help to me as I prepared this presentation of Holger Meyer, Paul Reimer, and Jorge Morfin of MIPP, E906, and MINERvA, respectively.

I thank each of them for their help. Of course, I have also drawn heavily on the presentations of Piermaria Oddone who has already produced a clear statement of the plans for Fermilab's future.